\documentclass[lettersize,journal]{IEEEtran}
\usepackage{amsmath,amsfonts}
\usepackage{array}
\usepackage[caption=false,font=normalsize,labelfont=sf,textfont=sf]{subfig}
\usepackage{textcomp}
\usepackage{stfloats}
\usepackage{url}
\usepackage{verbatim}
\usepackage{graphicx}
\usepackage[square,sort,comma,numbers]{natbib}
\def\BibTeX{{\rm B\kern-.05em{\sc i\kern-.025em b}\kern-.08em
    T\kern-.1667em\lower.7ex\hbox{E}\kern-.125emX}}
\usepackage{balance}
\usepackage{hyperref}       
\usepackage{amsmath}
\usepackage{amsfonts}
\usepackage{amssymb}
\usepackage{algorithm}
\usepackage{algpseudocode}
\usepackage{xcolor}
\usepackage{xspace}

\usepackage[textsize=footnotesize]{todonotes}
\usepackage{color}

\usepackage{comment}
\usepackage{algorithm}
\usepackage[algo2e,ruled,vlined,linesnumbered]{algorithm2e}
\usepackage{arydshln}
\usepackage{array}

\usepackage{booktabs}
\usepackage{acronym}
\usepackage{cleveref}
\newcommand{\myparagraph}[1]{~\\ \noindent \textbf{#1}.}
\newcommand{\vct}[1]{\boldsymbol{\ensuremath{#1}}}
\newcommand{\set}[1]{\ensuremath{\mathcal{#1}}}
\newcommand{\ef}{\ensuremath{\textit{ef}}\xspace}

\newcommand{\Dtot}{\ensuremath{\set{D}}\xspace}
\newcommand{\Dclean}{\ensuremath{\set{D}_c}\xspace}
\newcommand{\Dattk}{\ensuremath{\set{D}_p}\xspace}
\newcommand{\adv}{\ensuremath{\vct{\varepsilon}}}
\newcommand{\Dpoison}{\ensuremath{\set{D}^{\adv}_p}\xspace}

\newcommand{\C}{\ensuremath{\set{C}}\xspace}

\newcommand{\npoison}{\textit{s}}
\newcommand{\ntot}{\textit{n}}
\newcommand{\sonic}{\texttt{Sonic}\xspace}

\newcommand{\slowPoison}{\texttt{SlowP}\xspace}

\newcommand{\fishhdb}{\small F\ensuremath{_{\rm{\footnotesize HDBSCAN}}} \normalsize}
\newcommand{\fishdb}{\small F\ensuremath{_{\rm{\footnotesize DBSCAN}}}\normalsize}
\newcommand{\fishhier}{\small F\ensuremath{_{\rm{\footnotesize H_S}}}\normalsize}

\SetKwInput{KwInput}{Input}                
\SetKwInput{KwOutput}{Output}
\SetKwComment{tcp}{$\triangleright$\ }{}

\newcommand{\hdbscan}{HDBSCAN*\xspace}
\newcommand{\minpts}{\textit{minPts}\xspace}
\newcommand{\eps}{\ensuremath{\epsilon}\xspace}

\acrodef{HNSW}{Hierarchical Navigable Small World}

\newcommand{\updateAndCluster}{\texttt{update\_cluster}\xspace}
\newcommand{\M}{\textcolor{indianyellow}{$M$}}

\begin{document}
\title{\sonic: Fast and Transferable  Data Poisoning on Clustering Algorithms}

\author{
\IEEEauthorblockN{Francesco Villani\IEEEauthorrefmark{1}, Dario Lazzaro\IEEEauthorrefmark{2}\IEEEauthorrefmark{1}, Antonio Emanuele Cinà\IEEEauthorrefmark{1}, Matteo Dell'Amico\IEEEauthorrefmark{1}, Battista Biggio\IEEEauthorrefmark{3}, Fabio Roli\IEEEauthorrefmark{1}}\\
\vspace{0.05in}
\IEEEauthorblockA{\IEEEauthorrefmark{1} University of Genoa, Via Dodecaneso 35, Genoa, 16145, Italy}\\
\IEEEauthorblockA{\IEEEauthorrefmark{2} Università Sapienza di Roma, Via Ariosto 25, Rome, 00185, Italy}\\
\IEEEauthorblockA{\IEEEauthorrefmark{3} University of Cagliari, Via Marengo 3, Cagliari, 09100, Italy}
\vspace{0.05in}
}

\maketitle

\begin{abstract}
Data poisoning attacks on clustering algorithms have received limited attention, with existing methods struggling to scale efficiently as dataset sizes and feature counts increase. 
These attacks typically require re-clustering the entire dataset multiple times to generate predictions and assess the attacker's objectives, significantly hindering their scalability.
This paper addresses these limitations by proposing \sonic, a novel genetic data poisoning attack that leverages incremental and scalable clustering algorithms, e.g., FISHDBC, as surrogates to accelerate poisoning attacks against graph-based and density-based clustering methods, such as HDBSCAN.
We empirically demonstrate the effectiveness and efficiency of \sonic in poisoning the target clustering algorithms. 
We then conduct a comprehensive analysis of the factors affecting the scalability and transferability of poisoning attacks against clustering algorithms, and we conclude by examining the robustness of hyperparameters in our attack strategy \sonic.
\end{abstract}

\begin{IEEEkeywords}
data poisoning; adversarial machine learning; clustering robustness; transferability; machine learning security; unsupervised
\end{IEEEkeywords}

\section{Introduction}\label{sec:introduction}
Clustering algorithms are indispensable tools for organizing and deriving insights from vast amounts of unlabeled daily collected data~\cite{xu2015comprehensive, xu2005survey}.  
They facilitate the identification of meaningful patterns and structures by grouping similar data points into distinct clusters (i.e., groups), enabling valuable decision-making processes in various industries~\cite{ghosal2020short,ezugwu2022comprehensive},
including banking~\cite{bi2016big} and  healthcare~\cite{neijenhuijs2021symptom} among others.
In computer science, clustering algorithms are widely employed for tasks such as malware detection~\cite{rahat2024cimalir},
anomaly detection~\cite{chang2022video},
and topic modeling for identifying main themes in large text corpora~\cite{vayansky2020review}. 

However, the reliability and robustness of clustering algorithms 
have been questioned due to their proven vulnerability to data poisoning attacks~\cite{biggio2018wild,cina2023wild,cina2024machine}.
Data poisoning involves tampering with a small number of data points in the dataset, causing the clustering algorithm to produce inaccurate or misleading results, undermining the effectiveness and integrity of clustering algorithms. 
These attacks enable malicious operators to distort clustering outcomes by merging distinct clusters or generating practically unusable results. 
However, unlike the classification domain where much work has been developed~\cite{biggio2018wild, goodfellow2015explaining, croce2020robustbench, cina2024attackbench,cina2024sigma}, threats against unsupervised algorithms such as clustering are still little explored, along with the algorithms that can be used to test their robustness.
Additionally, developing attacks against clustering algorithms presents several challenges, such as the impossibility of exploiting fast gradient-based optimization algorithms~\cite{biggio2018wild,cina2024attackbench} or the computational complexity of the problem at hand~\cite{cina2022black}. 
Existing strategies~\cite{cina2022black,chhabra2020suspicion} necessitate re-clustering the entire dataset multiple times to optimize their attack to evaluate the candidate solutions. 
This process can quickly become impractical due to the substantial time and resource demands, particularly as datasets grow in the number of features and volume of samples. 
Consequently, benchmarking the robustness of clustering algorithms with existing tools becomes complicated, making it nearly impossible and exposing them to potentially malicious users. 

In this paper, we address this open challenge by leveraging two key observations: (i) data poisoning attacks often target a small percentage of data to preserve attack stealthiness and practicability~\cite{cina2023wild,carlini2023poisoning}; and (ii) the majority of clustering operations (e.g., distances between samples) remain valid for untainted data and do not require re-calculation.
From these two observations, we derive \sonic, a genetic optimization strategy to stage data poisoning attacks against clustering algorithms.
\sonic leverages a faster and incremental surrogate clustering algorithm to accelerate optimization, facilitating robustness verification testing procedures. 
We show that generated attacks from \sonic benefit of the transferability property, i.e., generated adversarial noise can successfully mislead the original target clustering algorithm.

Our experimental investigation provides compelling evidence of the remarkable performance of \sonic. 
We evaluate it on four benchmark datasets, including MNIST~\cite{LeCun2005TheMD}, FASHIONMNIST~\cite{xiao2017fashion}, CIFAR-10~\cite{krizhevsky2009learning}, and the 20 Newsgroups dataset~\cite{Lang95}. 
We compare its performance with state-of-the-art methodologies, showing that \sonic scales significantly better as both the number of samples and features in the dataset increase while preserving effectiveness, making it feasible to test the robustness of clustering algorithms on larger datasets. 
Lastly, we conduct a broader transferability analysis to evaluate the impact of \sonic perturbation on different clustering algorithms.
We conclude the paper by discussing related work, the main contributions, and future directions to address the limitations of the proposed approach.

The remainder of this paper is structured as follows: Section~\ref{sec:clustering_background} introduces the clustering algorithms under study and the surrogate model utilized. Section~\ref{sec:methodology} details our methodology, including the threat model considered in the paper, the problem statement, and the  \sonic data poisoning algorithm. 
Section~\ref{sec:exp_setup} describes the experimental setup, including the datasets and evaluation metrics adopted. 
Section~\ref{sec:exp_results} presents the experimental results, providing an in-depth analysis of \sonic's qualitative results and inner workings. 
Section~\ref{sec:related_works} reviews related work in unsupervised adversarial machine learning. 
We conclude the paper in Section~\ref{sec:conclusions} summarizing our contributions, key findings, and future directions to address the limitations of the proposed approach.

\section{Clustering Algorithms and Incremental Clustering}
\label{sec:clustering_background}
Data clustering is an unsupervised learning technique that creates groupings from unlabeled data; without access to label information, clustering algorithms are designed to uncover insights and identify shared patterns within the data itself.
Clustering algorithms can be classified into various categories depending on how they operate; prominent among these are hierarchical algorithms (e.g., hierarchical clustering using different linkages \cite{sneath2005numerical, ward1963hierarchical}) and density-based algorithms (e.g., DBSCAN~\cite{ester1996density}, OPTICS~\cite{ankerst1999optics}). Specifically, density-based algorithms are popular because they recognize clusters of arbitrary shape and have the feature of distinguishing noise points: items that are not assigned any cluster because they are in a low-density area~\cite{ester1996density}. Additionally, hybrid approaches, such as \hdbscan~\cite{campello_hdbscan}, combine features from multiple classes of algorithms.
In the following, we introduce the density-based, hierarchical, and hybrid clustering algorithms considered in this paper. Then, we present FISHDBC, an incremental algorithm used by the \sonic attack algorithm to accelerate the optimization process.


\subsection{DBSCAN and \hdbscan}
The seminal and most famous density-based clustering algorithm is DBSCAN~\cite{ester1996density}.
In density-based algorithms, data points with enough similar neighbors are considered to belong to \emph{dense areas} and will be assigned a cluster; the others are \emph{noise} and will be assigned to no cluster.
The similarity is generally expressed through a distance function; often, the samples to cluster are points in an Euclidean space, and the distance function chosen is Euclidean distance.
In DBSCAN, a data point is considered in a dense area and called \emph{core point} if at least \minpts points are at a distance at most \eps from it, where \minpts and \eps are algorithm parameters.
The algorithm then links in the same cluster the core points at a distance at most \eps from each other. 
Furthermore, DBSCAN has a concept of \emph{border points}, which are not in a dense area but within distance $\eps$ from points in a dense area themselves.
In an alternative implementation, DBSCAN*~\cite{campello_hdbscan},  border points are considered noise.
The time complexity and scalability of DBSCAN varies depending on dataset characteristics and parameter choices~\cite{gan2015dbscan,schubert2017dbscan}; in the worst case, it can grow up to \(O(\ntot^2)\).

\hdbscan~\cite{campello_hdbscan} is a popular evolution of DBSCAN; it can be seen as a variation that supports hierarchical clustering and uses a heuristic to detect appropriate values of \eps for different parts of the space to cluster.
As a result, the algorithm requires one less hard-to-tune parameter; additionally, unlike DBSCAN, it can recognize clusters of different densities within a single dataset.
Notably, we can observe parallelism between \hdbscan and hierarchical clustering. By setting $\minpts=1$, \hdbscan can effectively mimic the behavior of the single-linkage clustering algorithm.
Neither DBSCAN nor \hdbscan are incremental: if new elements are added to a dataset, the whole clustering has to be recomputed from scratch, requiring substantial resources.

\subsection{FISHDBC: Incremental Density-Based Clustering}

FISHDBC~\cite{dell2019fishdbc} is an algorithm that evolves \hdbscan in two directions: scalability for high-dimensional datasets or arbitrary distance/dissimilarity functions and support for incremental clustering.
It leverages \acp{HNSW}~\cite{malkov2020hnsw}, a data structure originally designed for approximate nearest-neighbor querying.
FISHDBC piggybacks on the distance computations carried out by \acp{HNSW} and uses them to maintain its data structures. 
As a result, FISHDBC approximates \hdbscan, and by tuning the \ef parameter, we can control the \ac{HNSW} search cost~\cite{dell2019fishdbc}. 
Specifically, by raising it, we increase the computational cost, and we obtain a closer approximation of \hdbscan.
Furthermore, an additional advantage of the algorithm's underlying structure is that FISHDBC is \emph{incremental}: once the initial clustering is computed, new elements can be added to the dataset, and the clustering can be updated with minimal computational effort.


Lastly, similarly to \hdbscan, FISHDBC offers the possibility of configuring the algorithm to approximate DBSCAN* and single-linkage clustering. When attacking those algorithms, FISHDBC is a particularly effective surrogate algorithm closely mimicking the algorithm under study.

\section{\sonic: Fast Data Poisoning Clustering}
\label{sec:methodology}

We present here \sonic, our genetic optimization strategy to rapidly stage poisoning attacks against clustering algorithms. 
In the following, we initially describe the threat model and then give a formal overview of the proposed attack, its algorithmic implementation, and convergence properties.

\subsection{Threat Model} Let $\Dtot = \{\vct{x}_i \in \mathbb{R}^d \}_{i=1}^{\ntot}$ and $\C: \mathbb{R}^{\ntot \times d} \rightarrow \mathbb{Z}^{\ntot}$ denote respectively the dataset to be poisoned by the attacker, where $\ntot$ is the number of samples and $d$ is the number of features, and the target clustering algorithm.
We assume the attacker aims to stage a data poisoning attack~\cite{cina2023wild} against the victim clustering algorithm $\C$, leading to incorrect groupings or misclassifications~\cite{cina2022black}.
To this end, we model a scenario where the attacker can only control a subset of samples $\Dattk \subseteq \Dtot$, with cardinality $\npoison$.
Realistically, the attacker has only a small percentage of control over the dataset because of access limitations or resource constraints~\cite{cina2023wild,shejwalkar2022back}, and does not influence remaining data $\Dclean$ ($\Dtot = \Dclean \cup \Dattk)$.
We indicate with $\Dpoison= \{\vct x_i + \vct z_i | \vct x_i \in \Dattk, \vct z_i \in \adv\}$ the corresponding data after the injection of the poisoning noise $\adv$ from the attacker to mislead data grouping from $\C$. 
Furthermore, following the principles in \cite{cina2022black} to make the attack more stealthy against human inspection of data, the attacker constrains the noise maximum intensity in $\adv$ and looks for the perturbation that minimizes the number of tampered features in data samples $\vct{x}_i$.
Unlike \cite{cina2022black}, we assume the attacker knows the internals of $\C$ and leverages such knowledge to make the attack faster on larger datasets.

\subsection{Problem Definition}
We define with $\C(\Dtot)$ the data grouping identified by $\C$ when applied to dataset $\Dtot$. 
The attacker aims to tamper the portion of data under their control, $\Dattk$, injecting an adversarial noise $\adv$ to make the clustering algorithm $\C$\ incorrectly group the data from the origin. This can be formalized as:

\begin{equation}
\begin{aligned}\label{eq:attack-objective}
    \min_{\adv \in \Delta} && \phi(\C(\Dtot), \C(\Dclean \cup \Dpoison) + \lambda \|\adv\|_0,\\ 
    \textrm{s.t.} &&  \Delta = \{\adv \in \mathbb{R}^{\npoison \times d}, \|\adv\|_{\infty} \leq \delta\} \, ,
\end{aligned}
\end{equation}

where $\phi$ is a similarity measure between clusterings~\cite{cina2022black} (e.g., AMI~\cite{AMI}, ARI~\cite{ARI}, or NMI~\cite{meilua2007comparing}). 
$\phi$ quantifies how well the clusters in one partitioning correspond to the clusters in another. 
By injecting the poisoned data $\Dpoison$, the attacker wishes to lower the similarity between clustering outcomes \C output on the untainted data \Dtot. 
The lower the score of $\phi$, the higher the success rate for the attacker.
Similarly to Cinà et al.~\cite{cina2022black}, we use a penalty term $\lambda \|\adv\|_0$ to the cost function to enforce the algorithm in search for poisoning samples with the minimal number of manipulated features. 
Lastly, the \emph{adversarial attack space} $\Delta$ defines the space of poisoning perturbation masks that satisfy the maximum power constraints of the attacker. 

\begin{algorithm*}[h!]
  \SetAlgoLined
\DontPrintSemicolon
\SetKwComment{tcp}{$\triangleright$\ }{}
\SetKwInput{KwInput}{Input}                
\SetKwInput{KwOutput}{Output}
\definecolor{indiagreen}{rgb}{0.07, 0.53, 0.03}
\definecolor{indianred}{rgb}{0.8, 0.36, 0.36}
\definecolor{ballblue}{rgb}{0.13, 0.67, 0.8}
\definecolor{indianyellow}{rgb}{0.89, 0.66, 0.34}

\SetCommentSty{mycommfont}

\KwInput{$\Dclean$, clean data; $\Dattk$, data under attacker's control; \C, target clustering algorithm; $\delta$, maximum intensity constraint; $\phi$, clustering similarity function; G, number of iterations; $\lambda$, Lagrangian multiplier.}
  \KwOutput{$\adv^\star$, optimal adversarial data poisoning noise.}
  
    $\vct \adv^{(0)} \gets \vct 0, \qquad \set{E} \gets \{\vct \adv^{(0)}\}$\label{alg:sonic-init}

     \M $\gets$ \texttt{prepare}(\C, \Dclean) \tcc*{Model based on the clean data \Dclean.} \label{alg:newmodel}

    \textcolor{indiagreen}{$\set{P}$} $\gets$ \updateAndCluster(\texttt{copy}(\M), $\Dattk$)\label{alg:sonic-ori-p} \tcc*{Clustering on $\Dtot=\Dclean \cup \Dattk$.}

    \For{g {\rm in} $1, \dots, G$} {

        \textcolor{indianred}{$\Dpoison$} $\gets \{x_i + \adv_i^{(g)} ~ | ~ x_i \in \Dattk, ~\adv_i^{(g)} \in \adv^{(g)}\}_{i=1}^{s}$\label{alg:sonic-dpois} \tcc*{Poison \Dattk.}
        
        \textcolor{indianred}{$\set{P}^{\adv}$} $\gets $ \updateAndCluster(\texttt{copy}(\M), \Dpoison) \label{alg:sonic-poison-p} \tcc*{Clustering on $\Dclean \cup \Dattk$.}
         
        $\Theta[\adv^{(g)}] \gets \phi($\textcolor{indiagreen}{$\set{P}$}, \textcolor{indianred}{$\set{P}^{\adv}$}) $+ \lambda \|\adv^{(g)}\|_0$\label{alg:sonic-eval} \tcc*{Score candidate $\adv^{(g)}$.}

        $\set{E} \gets \set{E} \cup \{\adv^{(g)}\}$ \label{alg:sonic-candidates} \tcc*{Increase population.}

        $\adv^{(g+1)}  \gets $ \rm \texttt{choice}($\Theta$)\label{alg:choice} 
        
        $\adv^{(g+1)}  \gets $ \rm \texttt{crossover}($\adv^{(g)}, \adv^{(g+1)}$)\label{alg:crossover} 
        
        $\adv^{(g+1)}  \gets $ \rm \texttt{mutation}($\adv^{(g+1)}, \delta$)\label{alg:mutation} 
    }
    \KwRet{$\adv^\star \in \arg\min\limits_{\adv \in \set{E}} \Theta(\adv)$}\label{line:end_attack}
    \caption{\sonic: Fast and Transferable Clustering Poisoning Attack}
\label{alg:sonic-pseudocode}
\end{algorithm*}

\subsection{Solution Algorithm}\label{sec:solution-alg}
We present here our attacking algorithm, \sonic, for solving \autoref{eq:attack-objective} . 
Taking inspiration from \cite{cina2022black}, we configure \sonic, depicted in \autoref{alg:sonic-pseudocode}, as a genetic black-box attack that, at each iteration, evaluates the adversarial noise offspring performance and produces new individuals with the crossover and mutation operators. 
After initializing the adversarial noise $\adv=\vct{0}$ and the population set $\mathcal{E}$ (\autoref{alg:sonic-init}), a new state $M$ 
based on the algorithm \C is created based on the clean data \Dclean (\autoref{alg:newmodel}).
The original clustering $\set{P}$ is computed by updating a copy of $M$ with \Dattk, and computing the clustering on the untainted data $\Dtot = \Dclean \cup \Dattk$ on the resulting model (\autoref{alg:sonic-ori-p}); we refer to this operation as \updateAndCluster.
From this point, the \sonic algorithm employs an iterative genetic approach to generate new offspring solutions $\adv$ to minimize the objective function specified in \autoref{eq:attack-objective}. 
In each iteration, \sonic performs the following steps: (i) creates a poisoned dataset $\Dpoison$ by injecting the current solution $\adv^{(g)}$ in $\Dattk$ (\autoref{alg:sonic-dpois}), (ii) applies our surrogate clustering algorithm after adding the poisoned data $\Dpoison$ to our model (\autoref{alg:sonic-poison-p}), (iii) evaluates the poisoning influence of $\adv^{(g)}$ based on the objective in \autoref{eq:attack-objective} (\autoref{alg:sonic-eval}), and updates the population set $\mathcal{E}$ (\autoref{alg:sonic-candidates}); (iv) the genetic operators \texttt{choice} (\autoref{alg:choice}), \texttt{crossover} (\autoref{alg:crossover}), and \texttt{mutation} (\autoref{alg:mutation}) are employed to produce the offspring adversarial noise for the subsequent iteration. 
Finally, \sonic returns the adversarial noise that minimizes \autoref{eq:attack-objective}, denoted as $\adv^\star$.
In the subsequent paragraphs, we expand on the implementation of the five key parts of \sonic, i.e., surrogate clustering (\autoref{alg:sonic-poison-p}), poisoning clustering evaluation (\autoref{alg:sonic-eval}), and the genetic operators (\autoref{alg:choice}-\autoref{alg:mutation}).

\myparagraph{Surrogate Clustering}
The main limitation encountered in state-of-the-art data poisoning attacks~\cite{cina2022black, chhabra2020suspicion} is their lack of scalability on large-scale data, as they require evaluating the target clustering algorithm \C on the entire dataset for each optimization iteration. 
Density-based clustering algorithms generally have non-trivial computation costs that depend on dataset characteristics, and this is exacerbated when, as in this case, clustering needs to be computed multiple times; in the worst case, computational complexity becomes \(O(\ntot^2)\)~\cite{gan2015dbscan,schubert2017dbscan}, with $\ntot$ being the cardinality of the whole dataset $\Dtot$.
However, we argue that the computational effort of these attacks can be substantially reduced by noting that data poisoning attacks typically target only a small percentage of the entire dataset~\cite{carlini2023poisoning, cina2024machine,cina2023wild}. 
It is thus reasonable to assume that most computations (e.g., distances between samples) remain valid for untainted data \Dclean and do not require significant adjustments or re-calculation.
\sonic leverages these observations, implementing them effectively in \autoref{alg:sonic-pseudocode}. 
Specifically, it uses incremental clustering algorithms (e.g., FISHDBC) as a \emph{surrogate algorithm} to decrease the computational costs of poisoning attacks, leveraging the above observations.
In \autoref{alg:newmodel}, \sonic prepares the clustering model $M$ on the clean dataset not controlled by the attacker, \Dclean. 
In this way, \sonic performs a partial computation based on the majority of input points that will not change and save its state.
Afterward, the final clustering for untainted data is obtained in \autoref{alg:sonic-ori-p} with the \updateAndCluster function, which updates the state $M$ with the data under the attacker's control, \Dattk, which are the only points that change during the optimization process. 
For each data sample in \Dattk, we add it to the saved state $M$ and update the clustering. 
The same principle is employed in \autoref{alg:sonic-eval}.
Rather than running the clustering algorithm \C on the whole dataset, as done in \cite{cina2022black}, only a portion of the data is now considered, leading to less demanding updates and thus drastically reducing the computational effort of \sonic.
Going into detail, as observed in later sections, we find that when the number of data points \(\npoison = \|\Dattk\|\) controlled by the attacker is small, the \sonic optimization procedure incurs drastically lower computational costs for the attacker. 
The time complexity for \updateAndCluster is, in practice, dominated mainly by \(\npoison\), which we can expect to be much smaller than $\ntot$~\cite{cina2023wild, carlini2023poisoning}. 
In summary, the higher efficiency \sonic offers relies on minimizing the recomputation required for untainted data and on fast incremental updates that mainly scale with respect to the number of poisoned samples.
Furthermore, we also demonstrate the significant effectiveness of \sonic in transferring the poisoning data, crafted with the incremental surrogate algorithm FISHDBC, on multiple target density-based and hierarchical clustering algorithms. 

\myparagraph{Poisoning Clustering Evaluation}
The $\phi$ function in \autoref{alg:sonic-eval} measures how the clusterings $\set{P}$ of \C on the untainted dataset $\Dtot$ differs from the groupings identified in $\set{P}^{\adv}$. 
Similarly to \cite{cina2022black}, we use the Adjusted Mutual Information (AMI) Score~\cite{AMI} as a similarity measure to evaluate the resulting clustering performance and to assess the impact of the adversarial perturbation. 
The AMI score between two clustering outcomes, $\set{P}$ and $\set{P}^{\adv}$, is formally defined as:
\begin{equation}
    AMI(\set{P},\set{P}^{\adv}) = {\frac{MI(\set{P}, \set{P}^{\adv}) - \mathbb{E}[MI(\set{P},\set{P}^{\adv})]}{\max{\{H(\set{P}), H(\set{P}^{\adv})\}} - \mathbb{E}[MI(\set{P}, \set{P}^{\adv})]}}
\end{equation}
where $MI(\set{P}, \set{P}^{\adv})$ represents the mutual information shared between the two clusterings. The term $\mathbb{E}[MI(\set{P},\set{P}^{\adv})]$ denote the expected mutual information.
The denominator's $\max{\{H(\set{P}), H(\set{P}^{\adv})\}}$ is the maximum of the entropies of the two clusterings, serving as an upper bound for $MI(\set{P}, \set{P}^{\adv})$.
AMI is equal to 1 when the two groupings $\set{P}$ and $\set{P}^{\adv}$ are identical, and 0 when they are independent of each other, i.e., they share no information.
The AMI score makes no assumptions about the cluster structure and performs well even in the presence of unbalanced clusters, a plausible scenario when the attacker stages a targeted attack by moving samples only from one cluster towards others~\cite{cina2022black}. 
Nevertheless, compared to the clustering algorithms considered by Cinà et al.~\cite{cina2022black}, density-based clustering algorithms, such as \hdbscan, support the notion of noise samples, i.e., points that do not fit well into any cluster and thus are considered anomalous. 
As a side effect of the AMI score, which looks at agreements between clustering outcomes, if two data points belonging to the same original cluster are marked as noise points after the poisoning process, the AMI score will be less affected.
To mitigate this issue, during evaluation, we assign a unique label to each noise sample to ensure that the presence of noise does not artificially inflate the AMI score. 
This change ensures that noise points do not contribute positively to the AMI score, making it more sensitive to actual clustering performance changes.

\myparagraph{Choice}
The selection operation in a genetic algorithm serves to choose individuals from the population to contribute to the next generation, giving preference to those with better fitness scores~\cite{eiben1991global}.
In \sonic (\autoref{alg:sonic-eval}), the choice operator applies a softmax function to the attacker's objective function (see \autoref{eq:attack-objective}) for each candidate in the population (i.e., $\Theta[\adv_{i}]$), assigning each a probability of being chosen.
Formally, the selection probability for a candidate $\adv_{i} \in \set{E}$ is inversely proportional to the value of \autoref{eq:attack-objective}, and is calculated as: 
\begin{equation}
    p( \adv_{i} ) = \frac{\exp(-\Theta[\adv_{i}])}{\sum_{\adv \in \set{E}} \exp(-\Theta[\adv])}\,
\end{equation}
Since the fitness score reflects both the attack's effectiveness and stealthiness, the selection process favors candidates that most effectively degrade the clustering results while maintaining minimal perturbation. 
Finally, \sonic employs an elitism strategy, maintaining a fixed population size and retaining only the most optimal offspring for the next generation.

\myparagraph{Crossover} 
The crossover operation merges genetic information from two parent individuals to generate one or more offspring~\cite{eiben1991global}. 
Crossover becomes fundamental in genetic algorithm for exploring new regions of the solution space and may reveal better performing candidates in future generations. 
Additionally, crossover helps to prevent the algorithm from prematurely converging on suboptimal solutions~\cite{eiben1991global}. 
In \sonic, we follow the crossover technique proposed in ~\cite{cina2022black}.
Specifically, the crossover is executed by blending the current candidate with another selected through the choice operation, with their components being randomly swapped with a probability $p_c$ (\autoref{alg:crossover}). 

\myparagraph{Mutation}
The mutation operation introduces random changes to individual genes within a candidate solution, thereby expanding the exploration of the solution space~\cite{eiben1991global}. 
Like the crossover operation, mutation's randomness helps prevent the algorithm from getting trapped in local optima by enabling the discovery of potentially more effective solutions. 
In \sonic, a candidate perturbation is mutated with a probability $p_m$ towards the nearest sample in the victim clusters, which encourages the merging of clusters. 
This strategy increases stealthiness by subtly shifting samples toward their neighbors and preserves effectiveness, as merging clusters significantly degrades clustering quality and, thus, the AMI score.
Higher values of $p_m$ enhance exploration, boosting the likelihood of finding diverse solutions, though this comes with increased stochasticity in the optimization process. 
Conversely, lower $p_m$ values lead to a more focused search, potentially accelerating convergence but with a higher risk of missing better solutions. 
Additionally, following the strategy in \cite{cina2022black}, a zeroing operation with a probability $p_z$ is used to eliminate perturbations on irrelevant features. This helps to refine the evolutionary process by concentrating the attack on more impactful changes. Combining mutation and zeroing ensures a balance between exploration and precision in crafting adversarial perturbations~\cite{cina2022black}.
\section{Experiments}\label{sec:experiments}
This section presents an overview of the experimental process and its results. 
We begin by detailing the experimental setup (Section~\ref{sec:exp_setup}), including a detailed description of the datasets used and any preprocessing steps applied to the data. We also explain the hyperparameter selection process in our experiments and outline the evaluation metrics employed to assess performance.
We then continue with the experimental results section (Section~\ref{sec:exp_results}), where we first evaluate \sonic's effectiveness in comparison to directly attacking \hdbscan, simulating attackers of varying strength. Next, we examine \sonic's scalability, emphasizing the benefits of using incremental clustering algorithms and testing different algorithm approximation levels to explore the trade-off between result quality and execution time for the attack.
We then analyze the transferability of perturbations generated by \sonic by applying them to different density-based and hierarchy-based clustering algorithms. Finally, we evaluate \sonic's empirical convergence properties and robustness to hyperparameter selection.

\subsection{Experimental Setup}\label{sec:exp_setup}
\myparagraph{Datasets}
We employ four well-known datasets: MNIST~\cite{LeCun2005TheMD}, FASHIONMNIST~\cite{xiao2017fashion}, CIFAR-10~\cite{krizhevsky2009learning}, and the 20 Newsgroups dataset~\cite{Lang95}. 
MNIST and FASHIONMNIST feature $28\times 28$ grayscale images of handwritten digits and fashion items, respectively.
The CIFAR-10 dataset contains color images of size $32\times 32$ pixels spanning 10 distinct classes.
We preprocess CIFAR-10 data by taking the embedding representation from the last convolutional layers of a pretrained ResNet-50\footnote{The model's architecture and weights have been sourced from the Torchvision library\cite{torchvision}.}~\cite{he2016deep} model trained on ImageNet. 
The 20 Newsgroups text dataset is a collection of approximately $20,000$ newsgroup documents partitioned across 20 different topics. 
We employ a three-step preprocessing approach to transform the textual data into a suitable format for the clustering algorithms under investigation. 
The first step involves processing the data~\cite{albishre2015effective}. We begin by removing links, special characters, and numbers. Furthermore, we tokenize the text, remove stopwords\footnote{Both the tokenizer and stopword remover are sourced from the nltk~\cite{bird2009natural} library.}, and convert the remaining tokens to lowercase. 
The next step involves using a pre-trained sentence transformer~\citep{reimers2019sentencebert} to convert the text documents into high-dimensional embeddings. 
Lastly, we apply Uniform Manifold Approximation and Projection (UMAP)~\citep{mcinnes2020umap} to reduce the dimensionality of the embeddings. We configure UMAP with $150$ components and $15$ neighbors and use cosine similarity as the distance metric for the dimensionality reduction process.

\myparagraph{Problem Setup}
Following the experimental setup of \cite{cina2022black}, we focus part of our analysis on binary clustering problems where samples belonging to a victim cluster are perturbed towards a target cluster.
Specifically, we consider samples from digits 0 and 4 for MNIST, labels `dress' and `ankle boot' for FASHIONMNIST, and the `automobile' and `frog' classes for CIFAR-10, respectively as, victim and target clusters.
We then follow the data extraction procedure from \cite{biggio2013data} and pick the $2,000$ closest samples to the centroids of the two distributions.
Complementary to our binary clustering analysis, we also consider a multi-cluster scenario. Specifically, for the 20 Newsgroups dataset, we select all samples belonging to four topics: `comp.windows.x', `rec.sport.baseball', `soc.religion.christian', and `sci.space', with a total of approximately 2,300 samples. 
Our goal is to attack data in `rec.sport.baseball' (victim cluster) and `soc.religion.christian' (target cluster).

\myparagraph{Evaluation Metrics}
The main properties of \sonic we want to evaluate are its effectiveness in disrupting the clustering process and the time it takes to execute the attack. To measure the effectiveness of the attack, we employ the AMI score between the clusterings generated from the clean and poisoned data, as defined in Section~\ref{sec:methodology}. Furthermore, to establish a baseline for comparison, we execute the attack without its incremental features by attacking directly \hdbscan, as done by Cinà et al.~\cite{cina2022black}. This allows us to evaluate and compare both attacks' performance in terms of results quality and attack efficiency.
All the experiments were run on a workstation equipped with an Intel(R) Xeon(R) Gold 5420 processor and 500GB of RAM.

\myparagraph{Clustering Setup}
Regarding the clustering algorithms\footnote{Sourced from Sklearn~\cite{scikit-learn}, Scipy~\cite{2020SciPy-NMeth}, and \hdbscan~\cite{mcinnes2017hdbscan} libraries.}, we adjust each algorithm's influential hyperparameters by conducting a grid search and selecting the configuration that guarantees the best results. Specifically, we tune hyperparameters such as $minPts$, minimum cluster size, and $\epsilon$ to obtain the best baseline groupings, using default values when possible. \citet{mcinnes2017hdbscan} provide an efficient Python implementation of \hdbscan which also supports computing DBSCAN*, the close relative of DBSCAN discussed in \Cref{sec:clustering_background}.
The FISHDBC implementation we use~\cite{fishdbc_github} is based on the \hdbscan implementation referenced above~\cite{mcinnes2017hdbscan}.

\myparagraph{Poisoning Setup}
We run \sonic with the penalty term set to $\lambda = 0.1$ and probabilities $p_c = 0.85$, $p_m = 0.15$, and $p_z = 0.05$. The total number of iterations is set to 110. Moreover, we set the \ef parameter of FISHDBC, used in the attack's optimization process, to $50$. Further considerations and analysis on \sonic's convergence properties and the influence of hyperparameters are provided in Section~\ref{sec:exp_results}. 
Lastly, we conduct a baseline comparison using the poisoning strategy outlined in Cinà et al.~\citet{cina2022black}, which involves using the target algorithm during the optimization process without any incremental approach. We refer to this attack as \slowPoison.

\subsection{Experimental Results}\label{sec:exp_results}

\begin{figure*}[t]
    \centering
    \includegraphics[width=0.49\textwidth]{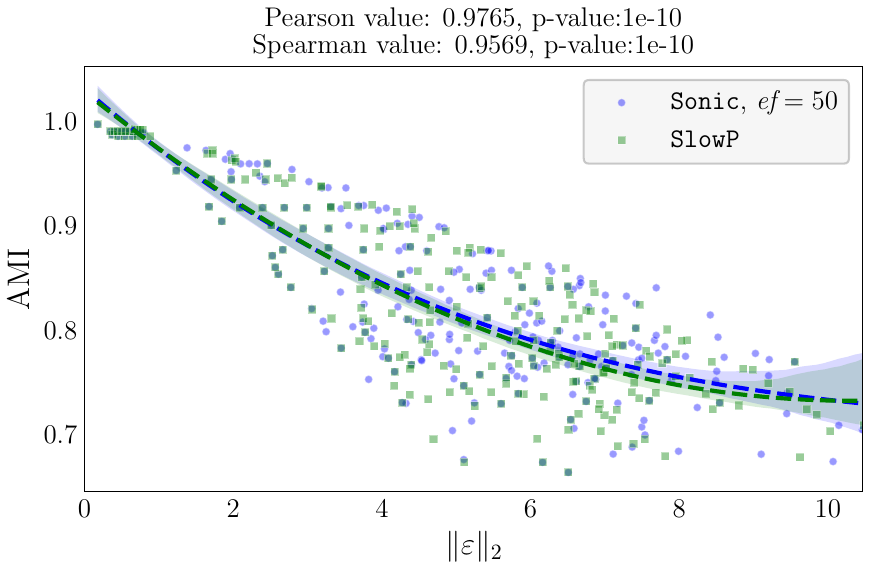}
    \includegraphics[width=0.49\textwidth]{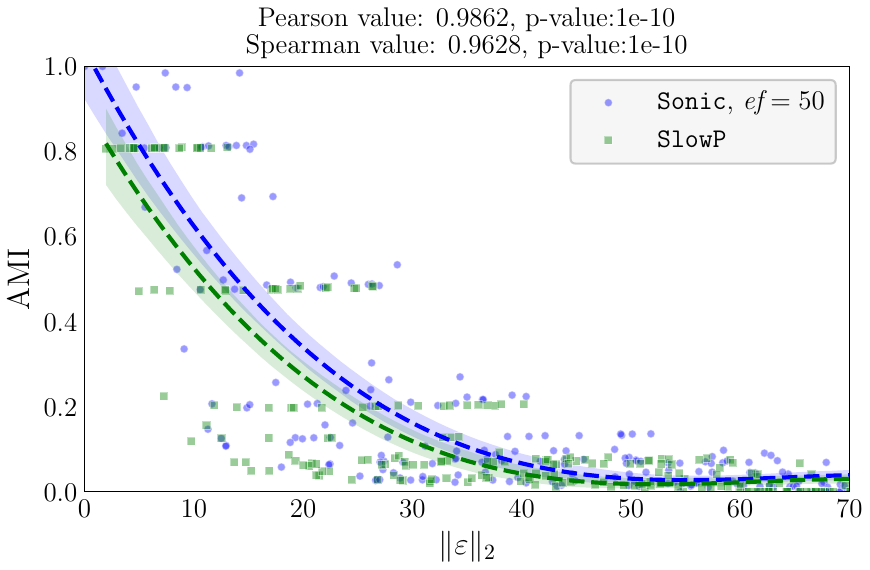}
    \includegraphics[width=0.49\textwidth]{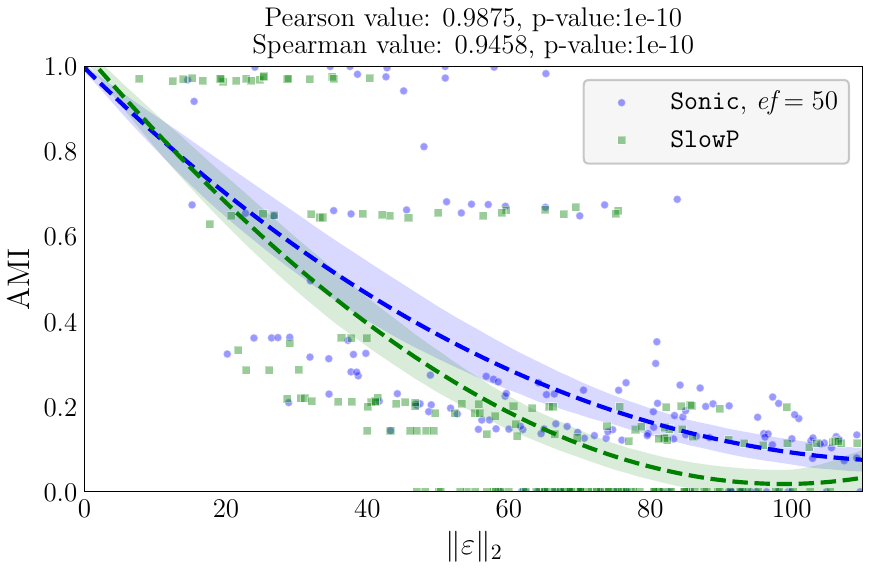}
    \includegraphics[width=0.49\textwidth]{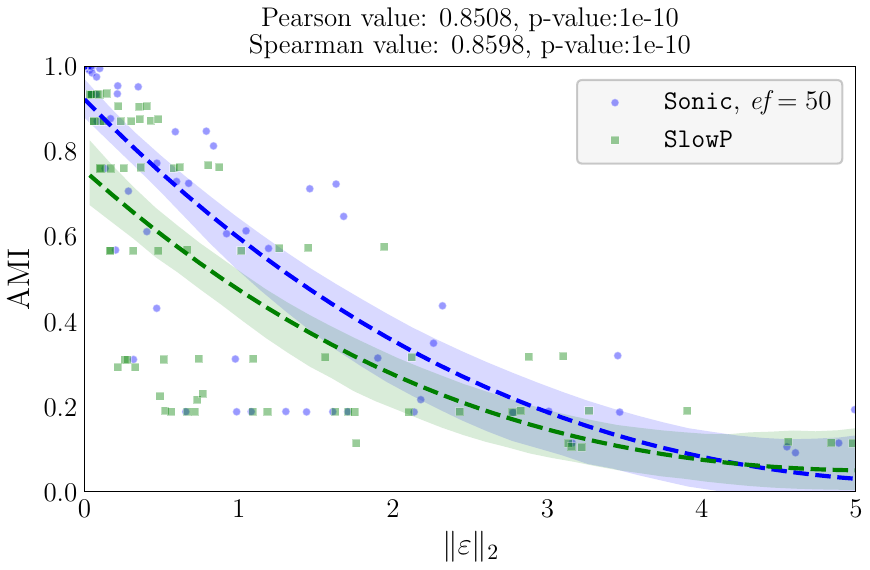}
    \caption{Robustness analysis is conducted on four datasets: 20 Newsgroups (top-left), MNIST (top-right), FASHIONMNIST (bottom-left), and CIFAR-10 (bottom-right). Each point in the plots represents the outcome of an $(s,\delta)$-experiment, where $s$ ranges from $0.01$ to $0.2$ and $\delta$ ranges from $0.05$ to $0.6$. A regression line is included to illustrate the trend of our results. Additionally, Pearson and Spearman values are reported to indicate the statistical significance of the correlation between the effectiveness of \sonic and \slowPoison.
    }
   \label{fig:robustness_analysis}
\end{figure*}

\myparagraph{Effectiveness}
We investigate whether \sonic can effectively disrupt the \hdbscan algorithm and compare its performance to the state-of-the-art method \slowPoison~\cite{cina2022black}. 
Our goal is to determine if \sonic, when combined with FISHDBC, can successfully transfer its data poisoning attack to the target algorithm.
To this end, we apply \sonic and \slowPoison to the four datasets, gradually increasing the perturbation size for each.
Specifically, for all datasets except CIFAR-10, we execute the attack using 20 different values for $s$ within the interval $[0.01, 0.2]$ and 12 values for $\delta$ within the interval $[0.05, 0.6]$, simulating attackers with varying strengths.
For CIFAR-10, due to the higher computational effort required to attack \hdbscan directly, we sample 9 values each from $[0.001, 0.1]$ for both $s$ and $\delta$.
Figure~\ref{fig:robustness_analysis} contains four scatterplots displaying the results of the attack procedures. Each point in the plots represents the outcome of a $(s, \delta)$-experiment, and the regression lines indicate the trends in the results. 
Firstly, we confirm the attack's correctness, as results progressively worsen with increasing perturbation magnitude.
Secondly, we identify a relationship between the two sets of results, represented in green (for \sonic) and blue (for \slowPoison), highlighted by the overlapping regression lines. 
Notably, we observe that the results for \sonic and \slowPoison are very similar to each other, and their correlation is further supported by the Pearson~\cite{pearson1895vii} and Spearman~\cite{spearman1961proof} correlation values displayed at the top of each plot. 
Ultimately, the results may differ across datasets or label configurations. On the 20 Newsgroups dataset, performance degradation is smooth, as both the $\delta$ and $s$ constraints influence the poisoning problem in a balanced way. 
This suggests that perturbing few samples with higher magnitude has the same effect of perturbing a higher number of them with decreasing intensity. 
In contrast, for other datasets, one constraint often dominates the other, leading to significant shifts in performance only when the more restrictive constraint is altered. 
For example, in the MNIST (top-right plot) and FASHIONMNIST (bottom-left plot) datasets, the parameter $s$ is the more influential factor, leading to a significant drop in the AMI score of the final clustering results. 
In other words, for these datasets, it is more effective for the attacker to perturb a larger number of data samples, even if the perturbations are less visible, rather than tampering with a few samples but with a high $\delta$.

\begin{figure*}[htb]
    \centering
    \includegraphics[width=0.49\textwidth]{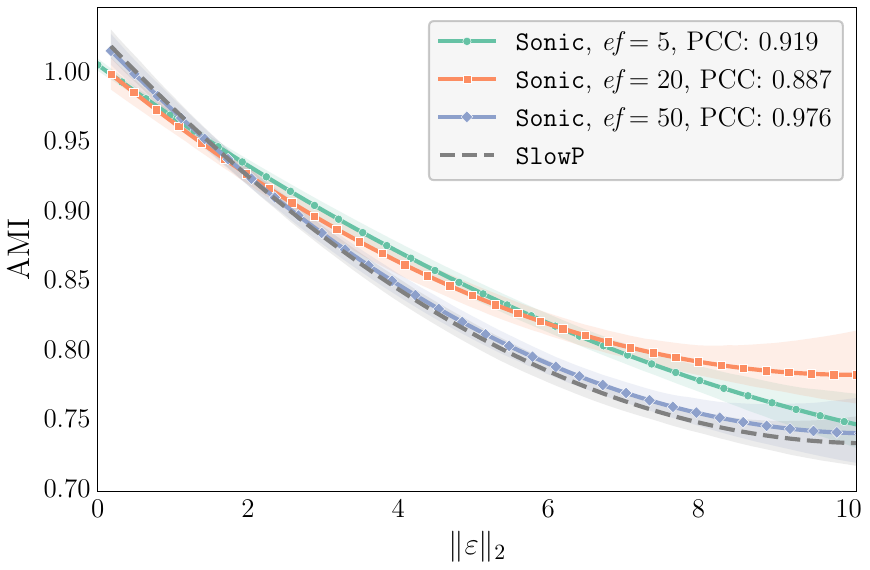}
    \includegraphics[width=0.49\textwidth]{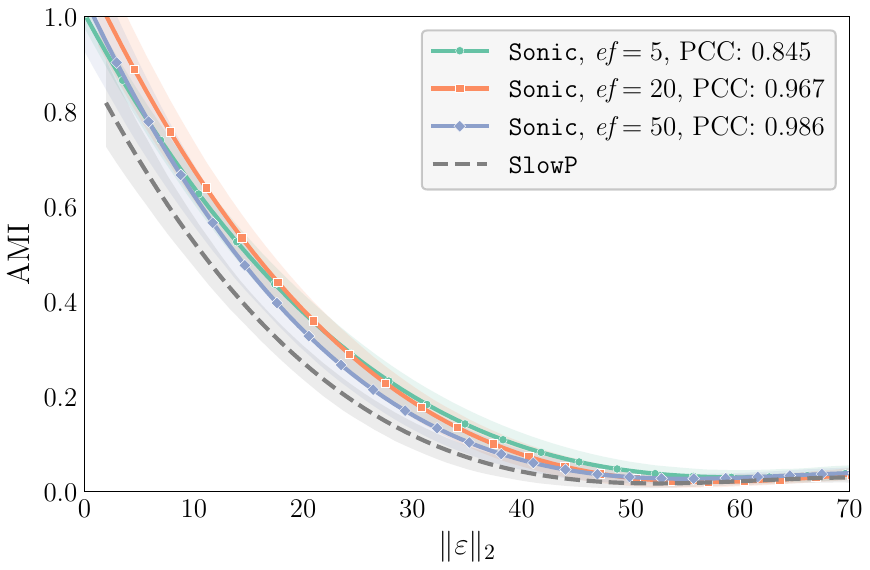}
    \includegraphics[width=0.49\textwidth]{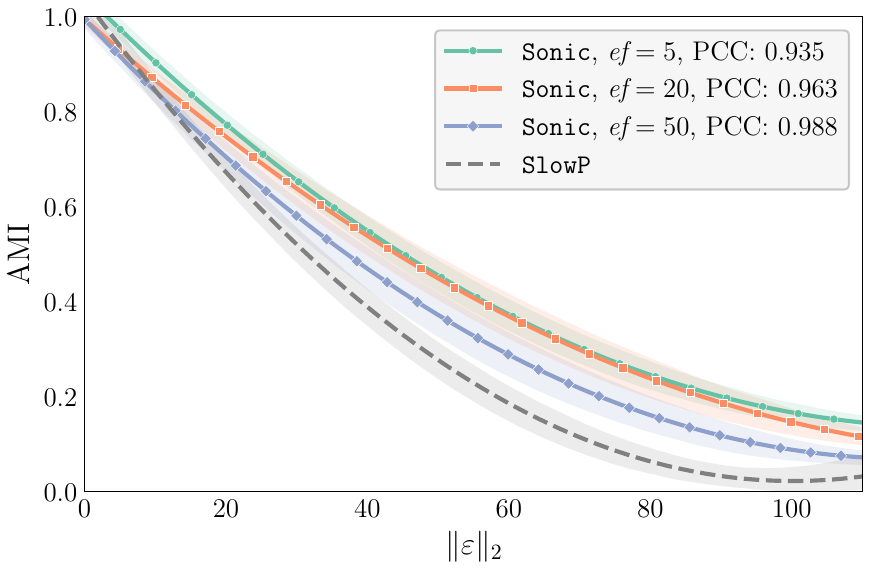}
    \includegraphics[width=0.49\textwidth]{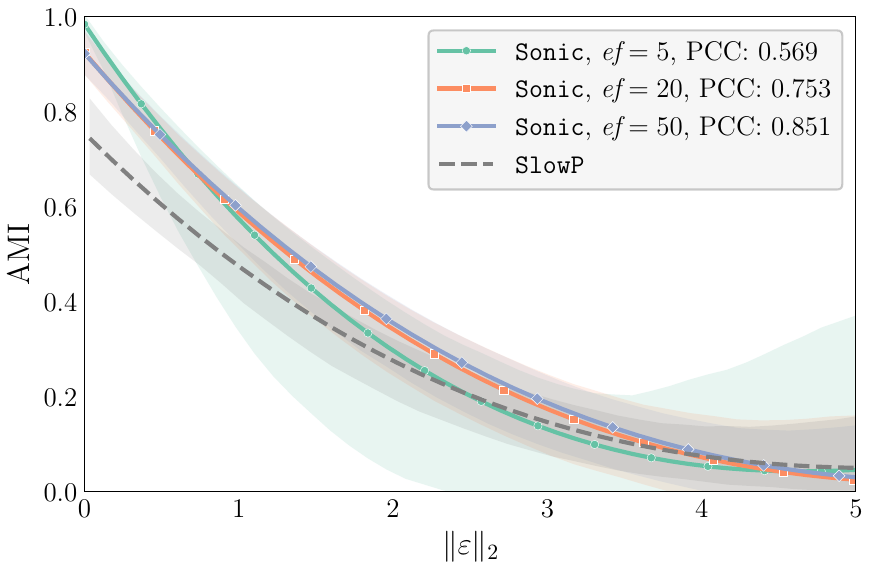}
    \caption{Robustness analysis on four datasets: 20 Newsgroups (top-left), MNIST (top-right), FASHIONMNIST (bottom-left), and CIFAR-10 (bottom-right). We present results for \sonic at different FISHDBC approximation levels (\ef), where lower \ef values indicate more accurate approximations of \hdbscan. The Pearson Correlation Coefficient (PCC) is provided to show the correlation between the effectiveness of \slowPoison and \sonic across various \ef levels.
}
    \label{fig:full_robustness_analysis}
\end{figure*}
We furthermore complete our effectiveness analysis by investigating the role of the \ef parameter of FISHDBC in \sonic.
We remark indeed that the \ef parameter controls the \ac{HNSW} search cost and the quality of the search process. 
The higher the \ef parameter, the more precise the approximation of FISHDBC toward the target \hdbscan clustering algorithm.
Figure~\ref{fig:full_robustness_analysis} presents regression lines for both \slowPoison and \sonic, configured with varying levels of FISHDBC approximation via the \ef parameter.
We also include the Pearson Correlation Coefficient (PCC) to assess the relationship between the results generated by different \sonic configurations and those produced by \slowPoison. 
As all the plots reveal, the regression lines generated with higher \ef values align more closely with that of \slowPoison, reflecting the improved accuracy of FISHDBC in approximating \hdbscan. 
Nevertheless, as we will see in the ``Efficiency" paragraph, higher accuracy levels affect the attack's execution time, highlighting the trade-off between precision and performance.

\begin{figure*}[htb]
    \centering
    \includegraphics[width=0.485\textwidth]{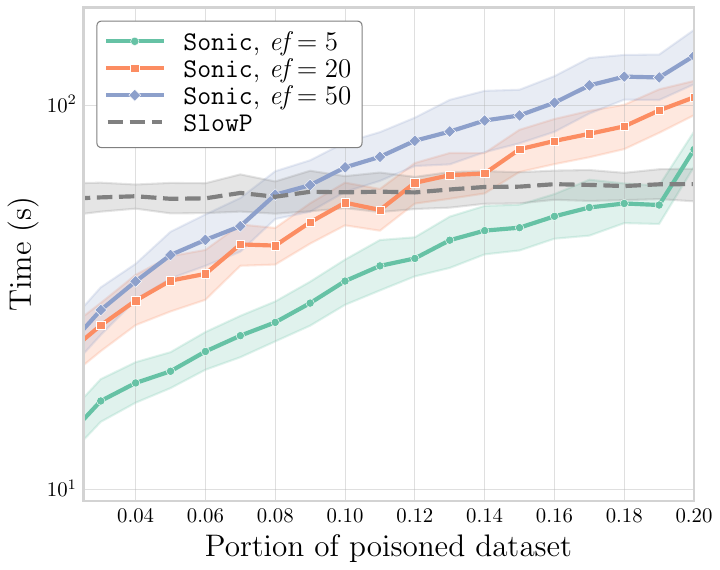}
    \includegraphics[width=0.485\textwidth]{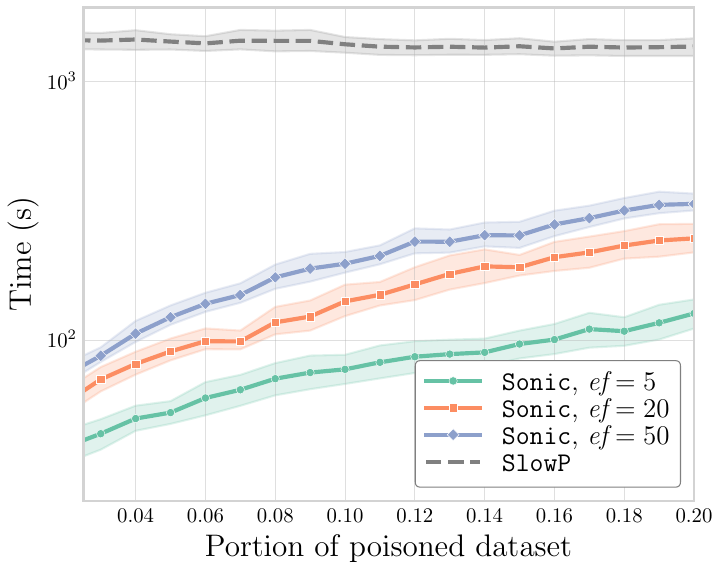} \hfill
    
    \includegraphics[width=0.485\textwidth]{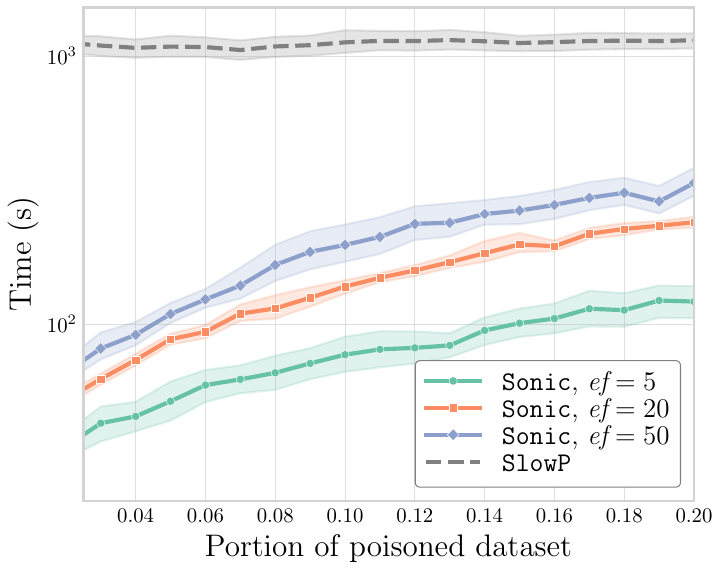}
    \includegraphics[width=0.485\textwidth]{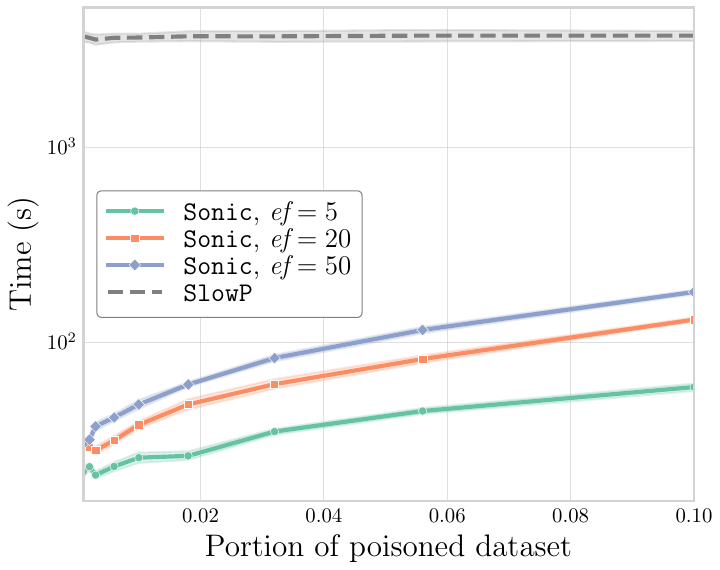}\hfill
    \caption{Time analysis for \sonic at different approximation levels (\ef) compared to \slowPoison on four datasets: 20 Newsgroups (top-left), MNIST (top-right), FASHIONMNIST (bottom-left), and CIFAR-10 (bottom-right). The x-axis represents the percentage of the dataset subjected to poisoning by the attacker, while the y-axis shows the total runtime of the attacks in seconds.}
    \label{fig:time_analysis}
\end{figure*}

\begin{figure*}[htb]
    \centering
    \includegraphics[width=0.485\textwidth]{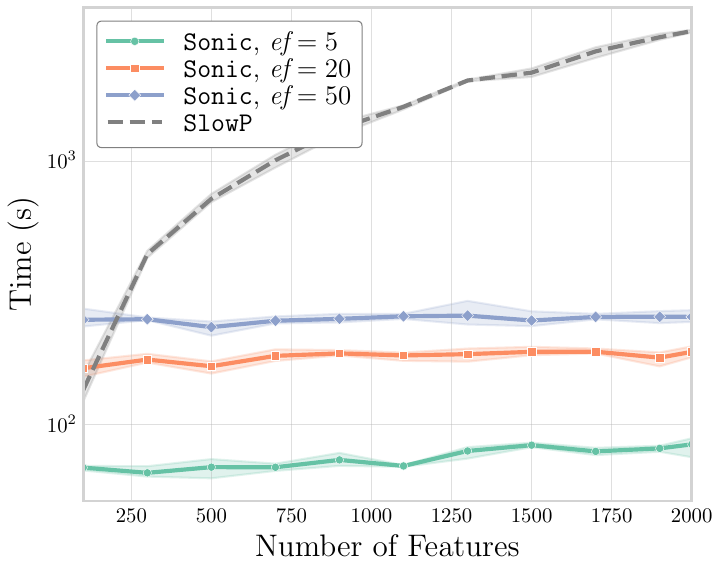}
    \includegraphics[width=0.485\textwidth]{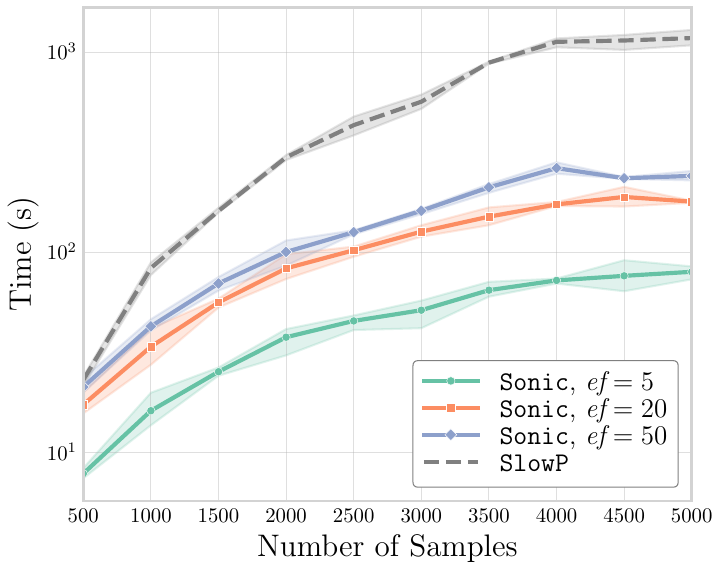}   \hfill 
    \caption{Scalability analysis of \sonic at different approximation levels (\ef) compared with \slowPoison. In the left plot, we increase the feature count of synthetic blob datasets, while in the right plot, we increase the number of samples, keeping the poisoning ratio fixed at $10\%$. The y-axis shows the total runtime of the attacks in seconds as the dataset dimensionality increases.}
    \label{fig:scalability_analysis}
\end{figure*}

\myparagraph{Efficiency}
We investigate the runtime execution of \sonic compared to \slowPoison to demonstrate \sonic's superior efficiency. 
Additionally, we examine the trade-off between \sonic's efficiency and the approximation quality, which is controlled by the \ef parameter. The results of these investigations are illustrated in Figures \ref{fig:time_analysis} and \ref{fig:scalability_analysis}. 
In more detail, similarly to the previous paragraph, Figure~\ref{fig:time_analysis} illustrates the execution time of the attack algorithms across the four datasets as the percentage of manipulated samples increases.
The percentage of manipulated samples, as depicted in the plots, plays a key role in determining the overall \sonic's execution time. 
This is a consequence of the fact that \sonic exploits the incremental nature of FISHDBC during the optimization of the poisoning samples. 
Particularly, \sonic updates the clustering algorithm computational statements only on the attacked points and not on the whole dataset (\autoref{alg:sonic-poison-p}). 
Consequently, the more points considered in the optimization process, the more statements \sonic will need to update.
When the number of manipulated data points is relatively low (a scenario commonly encountered in poisoning attacks~\cite{biggio2018wild,cina2023wild}) \sonic delivers outstanding performance gains, markedly reducing its execution time.
Conversely, for \slowPoison, the changing number of samples does not impact the final execution time; the algorithm maintains a consistent execution duration regardless of the number of perturbed data points. 
This is justified by the fact that \slowPoison updates the whole clustering outcomes at each attack iteration, hindering its practicability on large scale dataset. 
Going into the specifics of our results, we take into consideration, for example, the results for the MNIST dataset. 
Here \sonic with \ef set to $50$ and a poisoning ratio of $1\%$, is, on average, 27 times faster than \slowPoison, even when processing a modest set of $4,000$ samples.
When the poisoning ratio increases to $10\%$, \sonic remains $8$ to $10$ times faster than \slowPoison. %
These speed-ups are even more pronounced with the larger CIFAR-10 dataset, where \sonic on average is 84 times faster with 1\% poisoning ratio and 22 times faster when increasing the ratio to 10\%.
Finally, as shown in \autoref{fig:time_analysis}, these gaps can be adjusted by controlling the \ef parameter. 
Specifically, reducing the search cost of FISHDBC's \ac{HNSW} can indeed lower the quality of \sonic's results; however, it considerably shortens its execution time, offering additional flexibility when evaluating clustering algorithms' robustness.

\myparagraph{Scalability}
To build on the time analysis discussed previously, we now conduct a supplementary analysis focusing on scalability of \sonic compared to \slowPoison.
Specifically, we study the behavior of both attacks when dataset dimensionality (i.e., the number of samples and features) increases.
For this analysis, we utilize synthetic Gaussian blob datasets\footnote{The synthetic Gaussian blob datasets have been generated using Scikit-learn\cite{scikit-learn}}, varying in number of samples and number of dimensions. 
We first fix the number of samples at $4,000$, poisoning ratio to $10\%$, and vary the number of features within the range of $[100, 2.000]$. Subsequently, we set the number of features to $784$ and explore how performance changes with varying sample sizes, ranging from $500$ to $5.000$.
The results of our experiments are depicted in  Figure~\ref{fig:scalability_analysis}.
The plots show that the \sonic consistently has a lower execution time than \slowPoison, performing similarly or worse only when the dataset has really a few features (less than $200$) and samples (less than $500$). 
However, this phenomenon is not problematic since small datasets typically do not present scalability challenges.
When tackling larger datasets, \sonic scales efficiently with the dimensionality of the data, maintaining consistent execution times as the number of features increases. 
In contrast, \slowPoison experiences significant scaling issues as the number of features grows; furthermore, while both approaches experience increased computational times with a growing number of samples, \sonic scales better than \slowPoison.
Overall, \sonic leverages FISHDBC's underlying structure and its incremental design to more effectively manage large datasets, enabling the execution of poisoning attacks that would otherwise be unfeasible.

\myparagraph{Transferability}
We assess the \emph{transferability} impact of \sonic generated poisoned samples against various clustering algorithms. 
This property, well-known for supervised learning applications in adversarial machine learning~\cite{biggio2018wild,demontis2019adversarial}, is leveraged when the specific target model is unknown to the attacker.
In this context, the attacker creates attacks using surrogate models and then tests them on the target model. 
However, we are the first to study this property in the context of data poisoning attacks against clustering algorithms. 
The goal is to evaluate how effective these attacks are across different clustering algorithms, providing insights into their performance when the target model is not directly accessible. 
To achieve this, we generate poisoned data samples using \sonic with three distinct FISHDBC configurations to simulate the behavior of \hdbscan (denoted as \fishhdb), DBSCAN (denoted as \fishdb), and hierarchical single linkage (denoted as \fishhier).
Next, we inject these poisoned samples into the data and evaluate the performance of several target density-based and hierarchical clustering algorithms by running them on the poisoned dataset. 
Specifically, we test \hdbscan, DBSCAN, and hierarchical linkage methods, including single linkage (H$_{\text{S}}$), average linkage (H$_{\text{A}}$), complete linkage (H$_{\text{C}}$), and Ward's linkage (H$_{\text{W}}$). 
We explore two scenarios: a constrained scenario with low $\delta$ and $s$ values and a less constrained scenario with higher $\delta$ and $s$ values.
Tables~\ref{tab:transferability_mnist}-\ref{tab:transferability_news20} present the transferability experiments conducted on the MNIST and 20 Newsgroups datasets. 
The tables display the AMI scores obtained by testing the various clustering algorithms (rows) on poisoned datasets generated using the different configurations of \sonic (columns). 
The values in brackets represent the difference in AMI values between attacking the surrogate algorithm or directly the original, i.e., \sonic and \slowPoison.
As it can be seen in both tables, the poisoned samples generated using \sonic configuration with \fishhdb demonstrate the best overall transferability to all other algorithms under study. 
Additionally, the various hierarchical linkages, except for $H_S$, show greater robustness to the perturbations generated by both \sonic and \slowPoison. Nonetheless, \sonic configured with \fishhdb consistently produces results closest to the baselines.
\begin{table*}[htbp]
\caption{Transferability experiment results on the MNIST dataset. The columns indicate the clustering algorithm used as the source in \sonic to perform the poisoning attack, while the rows indicate the target clustering algorithms on which we test the transferability of the generated poisoned samples. The table presents two scenarios: the left side shows a low-budget scenario, representing a more constrained problem with $\delta = 0.2$ and $s = 0.01$, where the attacker can manipulate fewer samples with lower magnitude; the right side shows a high-budget scenario, with $\delta = 0.5$ and $s = 0.1$, allowing the attacker to manipulate a greater number of samples and features.}
\label{tab:transferability_mnist}
\centering
\setlength\tabcolsep{7.5pt}
\begin{tabular}{@{}llll@{\hspace{1.7cm}}lll@{}}
\toprule
 & \multicolumn{3}{c}{\textbf{Low Budget:} $\delta = 0.2$, $s=0.01$} & \multicolumn{3}{c}{\textbf{High Budget:} $\delta = 0.5$, $s=0.1$} \\
\cmidrule(r){2-4}\cmidrule(r){5-7}
 Target/Source & \fishhdb & \fishdb & \fishhier & \fishhdb & \fishdb & \fishhier \\ \midrule
 HDBSCAN & 0.15 (-0.03) & 0.38 (0.20) & 0.33 (0.16) & 0.00 (0.00) & 0.08 (0.08) & 0.10 (0.10) \\
 DBSCAN & 0.07 (0.00) & 0.07 (0.00) & 0.07 (0.00) & 0.07 (0.01) & 0.07 (0.01) & 0.07 (0.01) \\
H$_{\text{S}}$ & 0.04 (0.00) & 0.04 (0.00) & 0.04 (0.00) & 0.04 (0.00) & 0.04 (0.00) & 0.04 (0.00) \\
H$_{\text{C}}$ & 1.00 (0.02) & 1.00 (0.02) & 1.00 (0.02) & 0.69 (0.06) & 0.90 (0.27) & 0.94 (0.30) \\
H$_{\text{A}}$ & 0.99 (0.03) & 1.00 (0.04) & 1.00 (0.04) & 0.69 (0.03) & 0.85 (0.20) & 0.92 (0.27) \\
H$_{\text{W}}$ & 1.00 (0.02) & 1.00 (0.02) & 1.00 (0.02) & 0.69 (0.04) & 0.91 (0.26) & 0.94 (0.30) \\ \bottomrule
\end{tabular}
\end{table*}
\begin{table*}[!h]
\caption{Transferability experiment results on the 20 newsgroup dataset. The columns indicate the clustering algorithm used as the source in \sonic to perform the poisoning attack, while the rows indicate the target clustering algorithms on which we test the transferability of the generated poisoned samples. The table presents two scenarios: the left side shows a low-budget scenario, representing a more constrained problem with $\delta = 0.2$ and $s = 0.05$, where the attacker can manipulate fewer samples with lower magnitude; the right side shows a high-budget scenario, with $\delta = 0.5$ and $s = 0.1$, allowing the attacker to manipulate a greater number of samples and features.
}
\centering
\label{tab:transferability_news20}
\setlength\tabcolsep{7.5pt}
\begin{tabular}{@{}llll@{\hspace{1.7cm}}lll@{}}
\toprule
 & \multicolumn{3}{c}{\textbf{Low Budget:} $\delta = 0.2$, $s=0.05$} & \multicolumn{3}{c}{\textbf{High Budget:} $\delta = 0.5$, $s=0.1$} \\
\cmidrule(r){2-4}\cmidrule(r){5-7}
 Target/Source & \fishhdb & \fishdb & \fishhier & \fishhdb & \fishdb & \fishhier \\ \midrule
 HDBSCAN & 0.68 (-0.01) & 0.96 (0.27) & 0.96 (0.27) & 0.49 (0.00) & 0.96 (0.47) & 0.96 (0.47) \\
 DBSCAN & 0.20 (-0.01) & 0.21 (0.00) & 0.21 (0.00) & 0.20 (-0.02) & 0.21 (0.00) & 0.21 (0.00) \\
H$_{\text{S}}$ & 0.20 (-0.01) & 0.21 (0.00) & 0.21 (0.00) & 0.20 (-0.02) & 0.21 (0.00) & 0.21 (0.00) \\
H$_{\text{C}}$ & 0.86 (0.13) & 0.94 (0.20) & 0.94 (0.20) & 0.65 (-0.10) & 0.87 (0.13) & 0.87 (0.13) \\
H$_{\text{A}}$ & 0.77 (-0.01) & 0.94 (0.15) & 0.94 (0.15) & 0.63 (0.06) & 0.94 (0.37) & 0.94 (0.37) \\
H$_{\text{W}}$ & 0.96 (0.24) & 0.96 (0.24) & 0.96 (0.24) & 0.93 (0.35) & 0.97 (0.39) & 0.97 (0.39) \\ \bottomrule
\end{tabular}
\end{table*}

\begin{figure*}[htbp]
    \centering
    \includegraphics[width=0.495\textwidth]{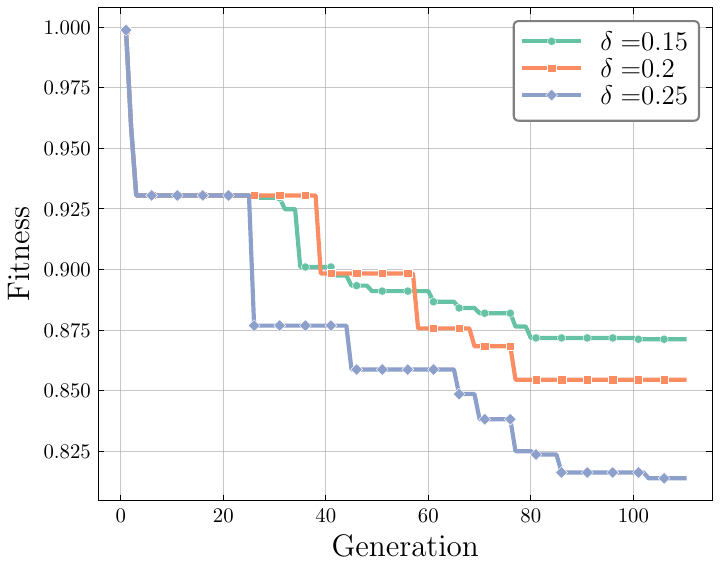}
    \includegraphics[width=0.495\textwidth]{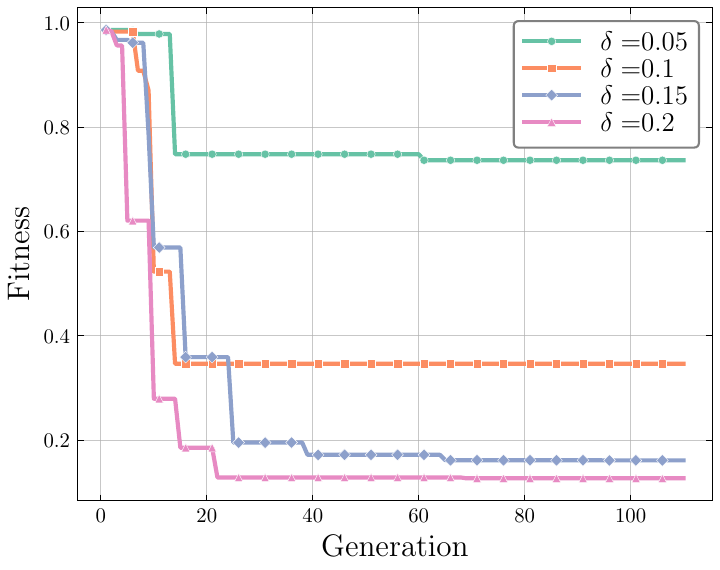}
\caption{Convergence curves of \sonic showing the best fitness value at each iteration. The left plot illustrates an example of convergence on the 20 Newsgroups dataset, with $\delta$ in $[0.15,0.25]$, while the right plot shows the convergence on the FASHIONMNIST dataset, with $\delta$ in $[0.05,0.2]$. The attacks have been run for 110 iterations each, fixing the poisoning ratio to $15\%$.} 
    \label{fig:convergence}
\end{figure*}

\myparagraph{Convergence}
We now examine the empirical convergence of \sonic, focusing on the number of iterations required for the algorithm to reach an optimum. Figure~\ref{fig:convergence} illustrates various constrained scenarios across two datasets, 20 Newsgroups and MNIST, highlighting the best fitness value achieved at each step of the iterative process. Our results demonstrate that \sonic converges within a relatively small number of iterations, consistently improving the best solution throughout the optimization. Although different constraints affect the optimal solution, all experimental configurations converge in approximately 110 iterations. 
%

\begin{figure*}[htbp]
    \centering

    \includegraphics[width=0.495\textwidth]{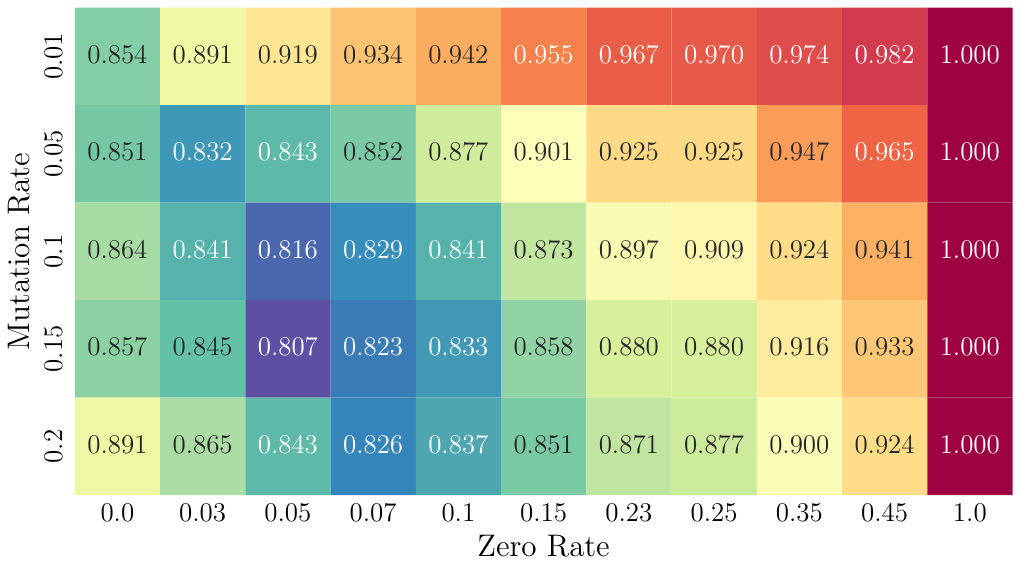}
    \hfill
    \includegraphics[width=0.495\textwidth]{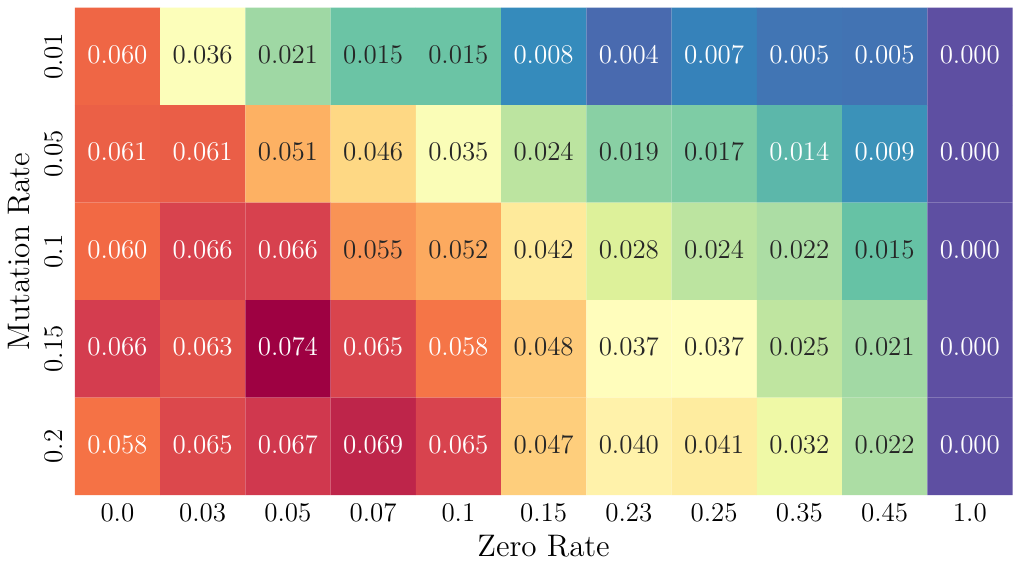}

    \includegraphics[width=0.495\textwidth]{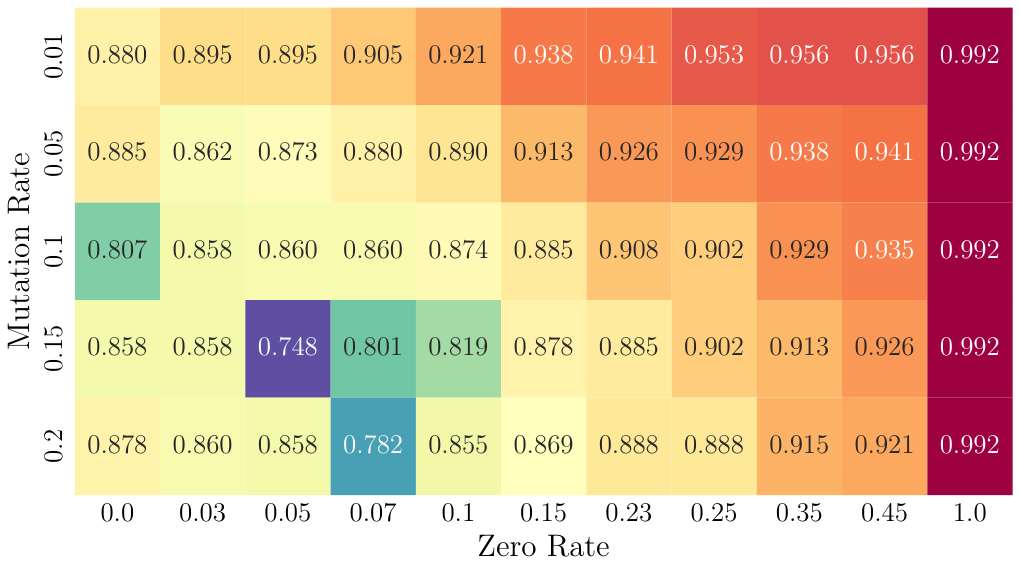}
    \hfill
    \includegraphics[width=0.495\textwidth]{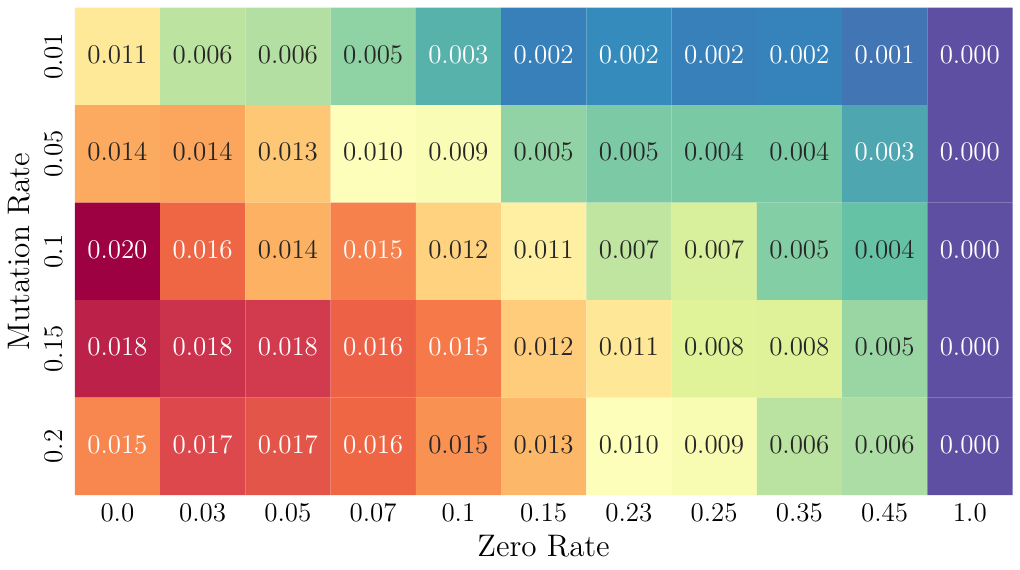}
    
    \caption{Hyper-parameter study for \sonic focusing on mutation probability ($p_m$) and zero probability ($p_z$) used in the mutation operator (presented in Section~\ref{sec:solution-alg}). The analysis is conducted on 20 Newsgroups (top row) with $s = 0.1$ and $\delta = 0.5$, and MNIST (bottom row) with $s = 0.05$ and $\delta = 0.2$. The left column displays AMI scores, while the right column shows the values of the cost function from~\autoref{eq:attack-objective}. The crossover probability ($p_c$) is fixed at $0.85$.}
    \label{fig:hyperparameter_heatmaps}
\end{figure*}

\myparagraph{Hyperparameters Study}
Lastly, we conduct a hyperparameter study on the mutation probability parameter $p_m$ and the zero mutation probability $p_z$ of \sonic, with the crossover probability fixed at $p_c = 0.85$. The attack is performed with different hyperparameter configurations, where $p_m$ ranges from $[0.01, 0.2]$ and $p_z$ ranges from $[0, 1]$. 
The results obtained by running \sonic with these configurations are shown in Figure~\ref{fig:hyperparameter_heatmaps}.
We perform the study in two different settings: the first on the multi-cluster problem using the 20 Newsgroups dataset and the second on the two-cluster problem using the MNIST dataset.
The heatmaps reveal that high values of the zero mutation probability $p_z$ significantly reduce noise and diminish the attack's effectiveness. Conversely, setting $p_z = 0$ results in more detectable perturbations and overall poorer performance. As noted in \citet{cina2022black}, maintaining low non-zero values of $p_z$ facilitates the optimization process in discovering better solutions while keeping the perturbation stealthier.
The mutation probability parameter $p_m$ shows resilience across different settings; nevertheless, very high $p_m$ values introduce greater stochasticity, which may increase the number of iterations needed to achieve optimal results.
Overall, the hyperparameters chosen in Section~\ref{sec:exp_setup} provide a balanced trade-off between stealthiness and effectiveness. This balance can be adjusted in practical scenarios according to the attacker's specific goals.

%


\section{Related Works}\label{sec:related_works}

The focus on robustness and adversarial machine learning has traditionally been on supervised learning, leaving unsupervised learning comparatively under-explored in this context. Initial contributions in this area were made by Dutrisac and Skillicorn~\cite{dutrisac2008hiding, skillicorn2009adversarial}, followed by a more comprehensive investigation by \citet{biggio2013data}, who proposed a categorization and theoretical framework for the challenge of adversarial clustering.
Additionally, the same work~\cite{biggio2013data} introduced a perfect knowledge attack targeting single-linkage clustering and, in a subsequent study, extended this to complete linkage hierarchical clustering~\cite{biggio2014poisoning}, particularly within the computer security domain.
\citet{crussell2015attacking} devised an attack on DBSCAN \citep{ester1996density} by exploiting the inherent vulnerability of density-based clustering algorithms, creating bridges between clusters. \citet{chhabra2020suspicion} relaxed the assumptions of perfect knowledge made by previous works by developing a derivative-free strategy that perturbs a single data point in a linearly separable task without requiring prior knowledge of the clustering algorithm's metric. This threat model is further addressed by \citet{cina2022black}, who designed a black-box poisoning attack using the Abstract Genetic Algorithm \citep{eiben1991global} framework. Their approach requires no knowledge of the clustering algorithm or its parameters and allows for the perturbation of multiple data points simultaneously. 
Further work has been conducted by \citet{xu2023a2sc}, where they propose an adversarial attack to fool subspace clustering. It achieves misclassification by applying adversarial manipulations inside the linear subspace to move a sample toward the target class.
In a recent study by \citet{zhang2024novel}, a new data poisoning attack was introduced, targeting deep clustering models such as deep k-means\citep{fard2020deep} and VaDE~\citep{jiang2017variational}, exploiting the robust features of clustering categories. Additionally, the study discusses creating adversarial examples against K-means and Gaussian mixture models by manipulating clean input to the decision boundary.

Overall scalability remains a primary concern when dealing with attacks against clustering algorithms~\cite{cina2022black,biggio2014poisoning,biggio2013data,zhang2024novel}, especially density and hierarchy-based. This challenge either slows down the algorithms or forces them to adopt heuristics to maintain efficiency.
\sonic addresses these issues by combining efficiency and efficacy, providing fast benchmarking of unsupervised systems while ensuring the quality of the resulting perturbations.
\section{Conclusions, Limitations, and Future Works}\label{sec:conclusions}
This work proposes \sonic, a genetic optimization poisoning attack against clustering algorithms. 
\sonic speeds up the poisoning process by acknowledging two main insights: (i) during practical poisoning attacks, only a small subset of data is tampered with by the attacker, and (ii) most of the clustering operations (e.g., distances between samples) on clean data points does not need to be re-computed. 
To this end, \sonic leverages a surrogate incremental clustering model, i.e., FISHDBC, to mitigate the scalability problems of previous state-of-the-art iterative methods, removing the burden of re-clustering the whole dataset at each optimization step. 
We report an experimental evaluation spanning four different datasets, both in the image and text domain, showing the effectiveness and efficiency of the attack. \sonic successfully disrupts target clustering algorithms, achieving comparable performance to attacks that directly incorporate the target algorithm in the optimization process at a fraction of the execution time.
Furthermore, we demonstrate that attacks generated with \sonic can effectively transfer to other clustering algorithms such as \hdbscan, DBSCAN, and hierarchical single linkage. However, methods like complete linkage and average linkage clustering exhibit greater resilience, highlighting the need for algorithm-specific attack strategies to better evaluate and enhance the robustness of various clustering techniques.
In conclusion, \sonic represents a significant advancement in robustness verification of clustering algorithms, enabling the benchmarking of unsupervised systems even as dataset sizes continue to grow.


\section*{Acknowledgments}
\label{sec:ack}
This work has been partially supported by the European Union's Horizon Europe research and innovation program under the project ELSA, grant agreement No 101070617;
by Fondazione di Sardegna under the project ``TrustML: Towards Machine Learning that Humans Can Trust’’, CUP: F73C22001320007;
by EU - NGEU National Sustainable Mobility Center (CN00000023) Italian Ministry of University and Research Decree n. 1033—17/06/2022 (Spoke 10);
and by project SERICS (PE00000014) under the NRRP MUR program funded by the EU - NGEU.
Lastly, this work has been carried out while Dario Lazzaro was enrolled in the Italian National Doctorate on Artificial Intelligence run by Sapienza University of Rome in collaboration with University of Genoa.

\bibliographystyle{IEEEtranN}
\bibliography{main}

\end{document}